\documentclass[12pt,english,preprint, aps,nofootinbib]{revtex4-1}
\usepackage[T1]{fontenc}
\usepackage[latin9]{inputenc}
\usepackage{babel}
\usepackage{amsmath}
\usepackage[unicode=true,pdfusetitle,
 bookmarks=true,bookmarksnumbered=false,bookmarksopen=false,
 breaklinks=false,pdfborder={0 0 1},backref=false,colorlinks=false]
 {hyperref}

\makeatletter
\usepackage{etoolbox}
\usepackage{breqn}

\makeatletter
\let\cat@comma@active\@empty
\makeatother

\newcommand{\breqnoverloadothers}
{%
    \renewenvironment{equation}{\ignorespaces\begin{dmath}}{\end{dmath}\ignorespacesafterend}%
    \renewenvironment{equation*}{\ignorespaces\begin{dmath*}}{\end{dmath*}\ignorespacesafterend}%
    \renewenvironment{multline}{\ignorespaces\begin{dmath}}{\end{dmath}\ignorespacesafterend}%
    \renewenvironment{multline*}{\ignorespaces\begin{dmath*}}{\end{dmath*}\ignorespacesafterend}%

}

\newcommand\breqnundefineothers
{%
    \renewenvironment{equation}{}{}%
    \renewenvironment{equation*}{}{}%
    \renewenvironment{multline*}{}{}%

}

\AtBeginEnvironment{dmath}{\breqnundefineothers}
\AtBeginEnvironment{dmath*}{\breqnundefineothers}

\AtBeginEnvironment{dgroup}{\def\breqnundefineothers{}\breqnoverloadothers}
\AtBeginEnvironment{dgroup*}{\def\breqnundefineothers{}\breqnoverloadothers}

\newcommand\brwrap[3]{%
  \setbox0=\hbox{$#2$}
  \left#1\vbox to \the\ht0{\hbox to 0pt{}}\right.\kern-.2em
  \begingroup #2\endgroup\kern-.15em
  \left.\vbox to \the\ht0{\hbox to 0pt{}}\right#3
}

\makeatother

\begin{document}
\title{Post AdS/CFT}
\author{David A. Lowe}
\affiliation{Department of Physics, Brown University, Providence, RI 02912, USA}
\author{Larus Thorlacius}
\affiliation{Science Institute, University of Iceland, Dunhaga 3, 107 Reykjavik,
Iceland}
\begin{abstract}
The Hamiltonian governing the gravitational interaction of $N$ relativistic
particles in a four-dimensional anti-de Sitter background is derived
to leading order in Newton's constant. The resulting pairwise interactions,
combined with the confining nature of motion in anti-de Sitter spacetime,
are expected to lead to classical chaos. In the context of the AdS/CFT
correspondence, the emergence of a chaotic classical limit on the
gravity side has important implications for the dual three-dimensional
conformal field theory, including that the spectrum of conformal primary
operators at strong coupling should exhibit level repulsion in line
with the Wigner surmise.
\end{abstract}
\maketitle

\section{Introduction}

The study of the gravitational interaction of $N$ particles in asymptotically
flat spacetime has a long history, beginning with the post-Newtonian
approach of \citep{Einstein1938}, where a derivative/momentum expansion
is made. Later a post-Minkowskian expansion was developed where one
expands in powers of Newton's constant, but retains all orders in
momentum \citep{Schafer:1986aa,Ledvinka_2008}. Such an expansion
is, for instance, relevant for deriving gravitational wave emission
from coalescing binary black hole systems. In the present work we
consider the gravitational interaction of $N$ particles in an asymptotically
anti-de Sitter (AdS) spacetime, hence we formulate a post-AdS expansion. 

Our motivation is rather different from the asymptotically flat case,
as our ultimate goal is to explore features of the anti-de Sitter
spacetime/conformal field theory (AdS/CFT) correspondence that are
relevant to the black hole information problem. In earlier work \citep{Lowe:2022cne},
we framed the formation of a typical small AdS black hole in four-dimensional
anti-de Sitter spacetime as the gradual coalescence a cloud of collapsing
particles, that is well described when the particles are well-separated
using the holographic reconstruction methods of HKLL \citep{Hamilton:2005ju,Hamilton:2006az,Hamilton:2006fh,Kabat:2011rz}.
A small AdS black hole has a mass that is smaller than the characteristic
AdS mass scale and a finite lifetime due to black hole evaporation.
Much of the literature on AdS black holes instead considers the limit,
where the black hole mass is large compared to the AdS scale and a
high temperature limit of the canonical ensemble matches well with
the large mass limit of the microcanonical ensemble, but large AdS
black holes essentially behave as stable massive remnants in the context
of the black hole information problem. 

Within the setup of \citep{Lowe:2022cne}, the gravitational collapse
to a small AdS black hole is understood as eigenstate thermalization
on the space of small black hole states. This assumes that the evaluation
of semi-classical observables involves averaging over a dense population
of energy eigenstates within a narrow range of energies determined
by the finite black hole lifetime. The microscopic counting of states
for the initial cloud of particles agrees with the Bekenstein-Hawking
entropy (up to numerical factors of order 1) \citep{Zurek:1982zz,mukhanov}
so there is indeed a large number of CFT states (of order $e^{S}$)
that correspond to small AdS black holes. However a puzzle emerges
because perturbative excitations in AdS have energies (or equivalently
conformal dimensions) that are quantized in units of the inverse AdS
radius of curvature. If the pattern of the perturbative level spacing
persists in the interacting theory, these states would have enormous
degeneracies, and the conditions for eigenstate thermalization \citep{srednicki},
would not be satisfied. In order for eigenstate thermalization to
hold, the spectrum of quantum energy eigenstates must for the most
part be non-degenerate, leading to a quasi-continuum of energy levels.
While this assumption receives indirect support in the literature
for states that collapse to black holes \citep{Sekino:2008he} there
is surprisingly little direct evidence that this is the case. In the
present paper we consider the classical limit of a \emph{typical }scattering
state, relaxing the condition of black hole formation, and ask whether
such a state exhibits classical chaos. 

Due to the negative spacetime curvature of AdS, massive particles
will be confined to the interior region, and interaction energies
will always be finite. As such, one has infinite ``dwell'' time
and even relatively simple pairwise interactions are expected to lead
to classical chaos in systems with more than just a few particles
(see below for further discussion). The case of massless particles
is more delicate. Null geodesics reach null infinity in AdS at finite
affine parameter, indicating that massless particles can reach null
infinity and that additional boundary conditions must be provided.
Here we have in mind simple energy conserving Dirichlet boundary conditions,
with a flat metric on the conformal boundary. Massless particles will
then reflect off the boundary at infinity in a geodesic approximation.
While their interactions vanish at infinity, the time average of the
interaction energy is finite and non-vanishing. Again this is expected
to lead to classical chaos.\footnote{We note that the geodesic approximation fails before massless radiation
reaches null infinity due to the infinite gravitational redshift in
AdS spacetime.. A more complete treatment, would involve solving the
massless field equations and consider the reflection of wave packets
from the asymptotic region. We expect the time average of the interaction
energy to be non-vanishing and finite in this case as well.}

At this level of approximation, the Lyapunov time associated with
chaos appears to be very long, of order $R_{AdS}^{3}/l_{pl}^{2}$,
with particles needing many AdS crossing times to randomize their
momenta by factors of order one. At higher orders in the post-AdS
approximation, higher order interactions will appear which will only
decrease the estimate of the Lyapunov time. However, in galactic N-body
simulations \citep{Portegies_Zwart_2022} in the asymptotically flat
case, the Lyapunov time for multiparticle states tends to be dominated
not by averages of long-range interactions of many particles, but
by close approach of pairs of particles. We expect to see a similar
phenomenon in asymptotically AdS spacetime. In particular, for particles
that collapse to black holes, we expect a thermalization time on a
timescale set by the local physics of the black hole, perhaps of order
$M\,$$\log M$. 

Having established that the classical limit of the bulk dual of a
holographic conformal field theory is chaotic, one may then follow
the general logic of the Wigner surmise \citep{wigner1955,wigner},
that the full quantum system will exhibit level repulsion. We note
that this holds for any finite value of the Lyapunov time, so establishing
an upper bound is sufficient to make the argument. Hence, for $N$-particle
scattering, we may expect a quasi-continuum of states with level spacing
of order $e^{-S(E)}$ where $S$ is the microcanonical entropy of
the conformal field theory at energy $E$. As the microcanonical entropy
at a given fixed energy is finite in the three (or higher) dimensional
conformal field theories appearing in the AdS/CFT correspondence,
our argument yields a sharp prediction for the spectrum of conformal
field theories in dimensions three and higher with holographic duals,
that generic primary operators will experience level-repulsion and
develop a near-continuous spectrum. 

It is interesting to compare to the situation in asymptotically flat
spacetime. In this case, the dwell time of typical incoming clouds
of particles will be finite, and one expects ordinary perturbative
scattering, perhaps governed by underlying integrability, rather than
chaotic dynamics. Moreover, in the atypical sector of the space of
states that contains black holes, they may be dressed by infinitely
degenerate soft hair in the classical limit \citep{Haco_2018}. Having
a negative cosmological constant avoids both of these issues. The
soft hair is no longer present, as the spectrum of excitations becomes
discrete. Likewise in AdS one expects integrability will only apply
to certain special limits of amplitudes where the dwell time can be
made finite. Conversely, in AdS one can expect eigenstate thermalization
to work for typical states (of energies larger than the Planck mass)
whereas in asymptotically flat spacetime it may only work for very
special families of states. This suggests that AdS/CFT in the limit
of vanishing cosmological constant provides a regularized description
of dynamical black hole formation in quantum gravity, while that dynamics
may be much harder to access by considering the asymptotically flat
case directly.

In the following we briefly review the Hamiltonian approach to gravity
applied to an anti-de Sitter background, and then derive the Hamiltonian
for $N$-particles interacting at leading order in Newton's constant
and to all orders in momentum. We end with some brief conclusions.

\section{Bulk Dynamics}

Following the Hamiltonian formulation of \citep{Arnowitt:1959ah,Arnowitt:1959eec,Arnowitt:1960es,Arnowitt:1960zzc}
and its generalization to anti-de Sitter spacetime in \citep{ABBOTT198276},
we begin with the gravitational action with no matter sources,
\begin{equation}
S_{grav}=\frac{1}{16\pi G_{N}}\int d^{4}x\,g^{1/2}\left(\pi^{ij}\partial_{0}g_{ij}+N\left(R^{(3)}-2\Lambda+\frac{1}{2}\pi^{2}-g_{ik}g_{jl}\pi^{ij}\pi^{kl}\right)+2N_{i}\pi_{\,\,|j}^{ij}\right)\,,\label{eq:gravaction}
\end{equation}
where $G_{N}$ is Newton's constant, $R^{(3)}$ is the three-dimensional
Ricci scalar, $\Lambda$ is the cosmological constant and $\pi_{\,\,|j}^{ij}$
is the three-dimensional divergence of the conjugate momentum on a
spatial hypersurface. We also have the definitions\footnote{Our conventions differ from \citep{ABBOTT198276} by a factor of $g^{1/2}$
in the definition of the conjugate momentum variable. Here $\pi^{ij}$
is a tensor rather than a tensor density as in \citep{ABBOTT198276}.}
\begin{align}
N & =(-g^{00})^{-1/2},\qquad N_{i}=g_{0i}\,,\label{eq:Ndef}\\
g & =\det(g_{ij})\,,\qquad\pi=g_{ij}\pi^{ij}\,,\label{eq:gdef}\\
\pi^{ij} & =N\left(\Gamma_{kl}^{0}-g_{kl}g^{mn}\Gamma_{mn}^{0}\right)g^{ik}g^{jl}\,.\label{eq:pidef}
\end{align}
With matter present, the constraint equations become 

\begin{align}
g^{1/2}\left(R-2\Lambda+\frac{1}{2}\pi^{2}-g_{ik}g_{jl}\pi^{ij}\pi^{kl}\right) & =-16\pi G_{N}T_{\,0}^{0}\,,\label{eq:hamconstr}\\
g^{1/2}\pi_{\,\,|j}^{ij} & =8\pi G_{N}T_{\,0}^{i}\,,\label{eq:momconstr}
\end{align}
where $T_{\,0}^{0}$ and $T_{\,0}^{i}$ are the energy and momentum
densities, respectively. For point particle matter we have
\begin{align}
T_{\,0}^{0} & =-\sum_{A}\left(g^{ij}p_{Ai}p_{Aj}+m_{A}^{2}\right)^{1/2}\delta(\mathbf{x}-\mathbf{x}_{A})\,,\label{eq:energydensity}\\
T_{\,0}^{i} & =-\sum_{A}g^{ij}p_{Aj}\delta(\mathbf{x}-\mathbf{x}_{A})\,,\label{eq:momentumdensity}
\end{align}
where $A$ is summed over the different point particle sources (see
for example \citep{Arnowitt:1960zzc}). The three-dimensional Dirac
delta functions are defined so that $\int d^{3}xf(\mathbf{x})\delta(\mathbf{x}-\mathbf{x}_{A})=f(\mathbf{x}_{A})$
for a smooth scalar function $f$ on a spatial hypersurface. 

Our strategy will be to solve the Hamiltonian constraint \eqref{eq:hamconstr}
and the momentum constraints \eqref{eq:momconstr} order-by-order
in $G_{N}$ and express the leading order Hamiltonian for $N$ interacting
particles in terms of the particle positions and momenta along with
the propagating degrees of freedom of the gravitational field. We
adapt the gauge conditions of \citep{Ohta:1973je,Schafer:1986aa}
for four-dimensional asymptotically flat spacetime to an asymptotically
AdS background with $\Lambda=-3$ as follows,
\begin{align}
g_{ij} & =\frac{e^{4\phi}}{z^{2}}\delta_{ij}+h_{ij}^{TT}\,,\label{eq:metricpert}\\
\delta_{ij}\pi^{ij} & =0\,.\label{eq:tracepi}
\end{align}
Here $h_{ij}^{TT}$ is a transverse-traceless metric perturbation,
$\delta^{ij}h_{ij}^{TT}=0$, $\delta^{ij}\nabla_{i}h_{jk}^{TT}=0$,
where the covariant derivative $\nabla_{i}$ is with respect to the
background hypersurface metric $g_{ij}^{(0)}=\delta_{ij}/z^{2}$.
The trace part of the metric involves $\phi=\phi_{1}+\phi_{2}+\ldots$,
where the subscript denotes the order of the expansion in $G_{N}$. 

The conjugate momentum $\pi^{ij}$ may be decomposed as follows \citep{D_Hoker_1999},
\begin{equation}
\pi^{ij}=\pi_{TT}^{ij}+\nabla^{i}\pi_{T}^{j}+\nabla^{j}\pi_{T}^{i}+\left(\nabla^{i}\nabla^{j}-\frac{1}{3}g^{(0)\,ij}\nabla^{2}\right)\pi_{L}\,,\label{eq:conjmet}
\end{equation}
into a transverse-traceless symmetric tensor satisfying $\nabla_{j}\pi_{TT}^{ij}=0$
and $\delta_{ij}\pi_{TT}^{ij}=0$, a transverse vector satisfying
$\nabla_{j}\pi_{T}^{j}=0$, and a scalar mode $\pi_{L}$. In general,
such a decomposition also includes a trace part but this vanishes
due to the gauge condition \eqref{eq:tracepi}.

Inserting the coordinate conditions \eqref{eq:metricpert}, \eqref{eq:tracepi}
and the decomposition \eqref{eq:conjmet} on the left-hand side of
the Hamiltonian constraint \eqref{eq:hamconstr} and working to second
order in perturbations, we find after some algebra that
\begin{equation}
\begin{aligned}g^{1/2}\left(R-2\Lambda\right) & =\sqrt{g^{(0)}}\Big(24(\phi_{1}+4\phi_{1}^{2}+\phi_{2})+8\big(\nabla^{2}\phi_{1}+(\nabla\phi_{1})^{2}+2\phi_{1}\nabla^{2}\phi_{1}+\nabla^{2}\phi_{2}\big)\\
 & +4h_{ij}^{TT}\nabla^{i}\nabla^{j}\phi_{1}+h_{TT}^{ij}h_{ij}^{TT}+\frac{3}{4}\nabla^{i}h_{TT}^{jk}\nabla_{i}h_{jk}^{TT}-\frac{1}{2}\nabla^{i}h_{TT}^{jk}\nabla_{k}h_{ij}^{TT}+h_{TT}^{ij}\nabla^{2}h_{ij}^{TT}\Big)\,,
\end{aligned}
\label{eq:hamconstr2}
\end{equation}
and
\begin{equation}
g^{1/2}\left(\frac{1}{2}\pi^{2}-g_{ik}g_{jl}\pi^{ij}\pi^{kl}\right)=\sqrt{g^{(0)}}\left(-\pi_{TT}^{ij}\pi_{ij}^{TT}-2\pi_{TT}^{ij}\tilde{\pi}_{ij}-\tilde{\pi}^{ij}\tilde{\pi}_{ij}\right)\,,\label{eq:momconstr2}
\end{equation}
where indices are raised and lowered using the background hypersurface
metric and we have introduced the shorthand notation
\begin{equation}
\tilde{\pi}_{ij}=\nabla_{i}\pi_{j}^{T}+\nabla_{j}\pi_{i}^{T}+\left(\nabla_{i}\nabla_{j}-\frac{1}{3}g_{ij}^{(0)}\nabla^{2}\right)\pi_{L}\,.\label{eq:shorthand}
\end{equation}

For the purposes of the present paper it is sufficient to expand the
left-hand side of the momentum constraint \eqref{eq:momconstr} to
first order in perturbations, giving

\begin{equation}
g^{1/2}\pi_{\,\,|j}^{ij}=\sqrt{g^{(0)}}g^{(0)ij}\left[\left(\nabla^{2}-2\right)\pi_{j}^{T}+\frac{2}{3}\nabla_{j}\left(\nabla^{2}-3\right)\pi^{L}\right]\,,\label{eq:firstordermomconstr}
\end{equation}
where we have used the decomposition \eqref{eq:conjmet}.

\section{First Order Solution}

Our goal is to compute the Hamiltonian at linear order in $G_{N}$
but to arbitrary order in momenta. Our first task is to compute the
order $G_{N}$ contributions to $\phi$ and $\pi^{ij}$. We will assume
our initial state does not contain any transverse-traceless gravitational
modes at order $G_{N}^{0}$, but that these will subsequently be generated
due to radiative couplings. The general techniques we employ in this
section were developed in \citep{Allen:1986aa,Antoniadis:1986sb}
and later adapted to anti-de Sitter spacetime in \citep{D_Hoker_1999},
where various Euclidean AdS Green functions are computed. 

\subsection{Scalar perturbation}

We begin by solving the Hamiltonian constraint at first order in $G_{N}$,
which reduces to the following equation for the leading order perturbation
of the metric trace,
\begin{equation}
\left(\nabla^{2}-3\right)\phi_{1}(\mathbf{x})=-2\pi G_{N}\sum_{A}\left(z_{A}^{2}\delta^{ij}p_{Ai}p_{Aj}+m_{A}^{2}\right)^{1/2}z_{A}^{3}\,\delta(\mathbf{x}-\mathbf{x}_{A})\,.\label{eq:scalar}
\end{equation}
The solution is a sum over the point particle sources,
\begin{equation}
\phi_{1}(\mathbf{x})=\sum_{A}\phi_{1A}(\mathbf{x})=2\pi G_{N}\sum_{A}\left(z_{A}^{2}\delta^{ij}p_{Ai}p_{Aj}+m_{A}^{2}\right)^{1/2}G(\mathbf{x},\mathbf{x_{A}})\,,\label{eq:phi1}
\end{equation}
where the defining equation for the scalar Green function is 
\begin{equation}
\left(\nabla^{2}-3\right)G(\mathbf{x},\mathbf{x'})=-\frac{1}{\sqrt{g^{(0)}}}\delta(\mathbf{x}-\mathbf{x'})\,.\label{eq:greenfunctioneq}
\end{equation}
AdS symmetry implies the Green function can be expressed as a function
of the geodesic distance $\lambda(\mathbf{x},\mathbf{x'})$ between
its arguments, but for our purposes it is more convenient to work
with the so-called chordal distance,
\begin{equation}
u(\mathbf{x},\mathbf{x'})=\frac{1}{2zz'}\left(\left(x-x'\right)^{2}+\left(y-y'\right)^{2}+\left(z-z'\right)^{2}\right),\label{eq:chordaldistance}
\end{equation}
which is related to the geodesic distance by $u+1=\cosh\lambda$.
The chordal distance and its derivatives satisfy a number of useful
identities that are listed in \citep{D_Hoker_1999}. The ones that
enter into our considerations are included in Appendix \ref{sec:appA}
below for easy reference.

Inserting $G(\mathbf{x},\mathbf{x'})=G(u)$ into \eqref{eq:greenfunctioneq}
and using the identities \eqref{eq:id1} and \eqref{eq:id2}, the
equation for the scalar Green function reduces to
\begin{equation}
u(u+2)G''(u)+3(u\text{+1)}G'(u)-3G(u)=-z^{\prime\,3}\delta(\mathbf{x}-\mathbf{x'})\,.\label{eq:inhomeq}
\end{equation}
The corresponding homogeneous differential equation has the following
general solution,
\begin{equation}
G(u)=c_{1}(u+1)+c_{2}\frac{2u(u+2)+1}{\sqrt{u(u+2)}}\,.\label{eq:phi1general}
\end{equation}
Setting $c_{1}=-2c_{2}$ ensures maximal fall-off as $u\rightarrow\infty$.
If we further choose $c_{1}=\frac{1}{4\pi}$ we have in fact obtained
a correctly normalized solution to the full inhomogeneous problem
in \eqref{eq:inhomeq}.\footnote{This is easily verified by multiplying both sides of \ref{eq:inhomeq}
by a test function in $\mathbf{x}$ and integrating over a small spherical
volume centered on $\mathbf{x'}$.} The resulting scalar Green function is given by 
\begin{equation}
G(u)=\frac{1}{4\pi\sqrt{u(u+2)}}\left(\sqrt{u(u+2}-u-1\right)^{2},\label{eq:phi1sol}
\end{equation}
and the leading order perturbation of the metric trace is given by
the sum $\phi_{1}(\mathbf{x})=\sum_{A}\phi_{1A}(\mathbf{x})$, where
\begin{equation}
\phi_{1A}(\mathbf{x})=\frac{G_{N}}{2}\left(z_{A}^{2}\delta^{ij}p_{Ai}p_{Aj}+m_{A}^{2}\right)^{1/2}u_{A}^{-1/2}(u_{A}+2)^{-1/2}\left(\sqrt{u_{A}(u_{A}+2)}-u_{A}-1\right)^{2}\,,\label{eq:phi1final}
\end{equation}
with $u_{A}=u(\mathbf{x},\mathbf{x_{A}})$. We note that the metric
trace perturbation falls off as
\begin{equation}
\phi_{1}\sim z^{3}\label{eq:falloff}
\end{equation}
near the boundary at $z\to0$. This will be important when we consider
the boundary Hamiltonian in Section \eqref{sec:Hamiltonian}.

\subsection{Conjugate momentum perturbation}

Next we need the first order solution for the momentum constraints,
\begin{equation}
\left(\nabla^{2}-2\right)\pi_{i}^{T}(\mathbf{x})+\frac{2}{3}\nabla_{i}\left(\nabla^{2}-3\right)\pi^{L}(\mathbf{x})=-8\pi G_{N}\sum_{A}\frac{p_{Ai}}{\sqrt{g^{(0)}}}\delta(\mathbf{x}-\mathbf{x}_{A})\,.\label{eq:momentumone}
\end{equation}
 To solve this, we act with a transverse vector projector on the source
term to give the equation for the transverse vector Green function,
\begin{equation}
\left(\nabla^{2}-2\right)G_{ij'}^{T}(\mathbf{x},\mathbf{x'})=-\frac{g_{ij'}^{(0)}}{\sqrt{g^{(0)}}}\delta(\mathbf{x}-\mathbf{x'})+\nabla_{i}\frac{1}{\nabla^{2}}\nabla_{j'}\,,\label{eq:transgreen}
\end{equation}
which will generate the solution for $\pi_{i}^{T}$. The transverse
vector Green function is a bivector, with the unprimed index associated
to position $\mathbf{x}$ and the primed index associated to position
$\mathbf{x'}$. The parallel transporter, denoted by $g_{ij'}^{(0)}$,
maps the unit vectors tangent to the geodesic connecting $\mathbf{x}$
and $\mathbf{x'}$ into each other via the relation
\begin{equation}
t_{i}=-g_{\phantom{{(o)}}i}^{(0)\:j'}\,t_{j'}\,,\label{eq:transportrelation}
\end{equation}
where $t_{i}=\partial_{i}u/\sqrt{u(u+2)}$ and $t_{j'}=\partial_{j'}u/\sqrt{u(u+2)}$.
The parallel transporter can be expressed in terms of derivatives
of $u$ as follows \citep{D_Hoker_1999},\footnote{We note that the expression for $g_{ij'}^{(0)}$ in Table 2 of \citep{D_Hoker_1999}
contains a typo.}
\begin{equation}
g_{ij'}^{(0)}=-\partial_{i}\partial_{j'}u+\frac{\partial_{i}u\partial_{j'}u}{u+2}\,,\label{eq:paralleltransp}
\end{equation}
as can be verified by insertion into \eqref{eq:transportrelation}.
This expression will be useful momentarily, when we solve for $G_{ij'}^{T}(\mathbf{x},\mathbf{x'})$.

To generate the solution for $\pi_{L}$ we instead project onto the
longitudinal component of the source. This yields the Green function
equation
\begin{equation}
\frac{2}{3}\nabla_{i}\left(\nabla^{2}-3\right)G_{j'}^{L}=-\nabla_{i}\frac{1}{\nabla^{2}}\nabla_{j'}\,.\label{eq:greenfcneq}
\end{equation}
Let us consider \eqref{eq:greenfcneq} and \eqref{eq:transgreen}
in turn.

\subsubsection{Longitudinal component}

To proceed we need to solve for the inverse Laplacian $1/\nabla^{2}$
via
\begin{equation}
\nabla^{2}\Delta_{0}(\mathbf{x},\mathbf{x'})=-\frac{1}{\sqrt{g^{(0)}}}\delta(\mathbf{x}-\mathbf{x'})\,.\label{eq:inverselap0}
\end{equation}
Inserting $\Delta_{0}(\mathbf{x},\mathbf{x'})=\Delta_{0}(u)$ leads
to the following ordinary differential equation,
\begin{equation}
u(u+2)\Delta_{0}''(u)+3(u\text{+1)}\Delta_{0}'(u)=-z^{\prime\,3}\delta(\mathbf{x}-\mathbf{x'})\,.\label{eq:inhomeq2}
\end{equation}
The general solution to the homogenous problem is
\begin{equation}
\Delta_{0}(u)=c_{1}+c_{2}\frac{u+1}{\sqrt{u(u+2)}}\,.\label{eq:homsol}
\end{equation}
The coefficients $c_{1}$and $c_{2}$ are again determined by requiring
maximal fall-off at infinity and the correct normalization in the
$u\to0$ limit to match the $\delta(\mathbf{x}-\mathbf{x'}$) on the
right hand side of \eqref{eq:inverselap0},
\begin{equation}
\Delta_{0}(u)=\frac{1}{4\pi}\left(\frac{u+1}{\sqrt{u(u+2)}}-1\right)\,.\label{eq:inverselap1}
\end{equation}
 This can now be inserted on the right hand side of the equation for
the longitudinal Green function,
\begin{equation}
\frac{2}{3}\left(\nabla^{2}-3\right)G_{j'}^{L}=-\frac{1}{\nabla^{2}}\nabla_{j'}=\left(\nabla_{j'}\frac{1}{\nabla^{2}}\right)=-\left(\partial_{j'}u\right)\Delta_{0}'(u)\,.\label{eq:longeqn}
\end{equation}
We look for a solution of the form
\begin{equation}
G_{j'}^{L}(\mathbf{x},\mathbf{x'})=\left(\partial_{j'}u\right)a(u)\,,\label{eq:longsol}
\end{equation}
with $a(u)$ to be determined. Inserting this ansatz on the left hand
side of \eqref{eq:longeqn}, and using the identities \eqref{eq:id1}
- \eqref{eq:id4} in Appendix \eqref{sec:appA}, leads to an inhomogeneous
ordinary differential equation for $a(u)$,
\begin{equation}
u(u+2)a''(u)+5(u+1)a'(u)=\frac{3}{8\pi}\frac{1}{u^{3/2}(u+2)^{3/2}}\,,\label{eq:inhomogeneous}
\end{equation}
whose general solution is given by 
\begin{equation}
a(u)=c_{1}\frac{(u+1)(2u^{2}+4u-1)}{3u^{3/2}(u+2)^{3/2}}+c_{2}-\frac{2u^{2}+6u+3}{8\pi\sqrt{u}(u+2)^{3/2}}\,.\label{eq:longgensol}
\end{equation}
To ensure maximal falloff at large $u$ we get the condition $\frac{2}{3}c_{1}+c_{2}-\frac{1}{4\pi}=0$.
As $u\to0$ the solution goes like
\begin{equation}
a(u)=-\frac{c_{1}}{3\left(2u\right)^{3/2}}+O\left(\frac{1}{\sqrt{u}}\right)\,,\label{eq:u0limit}
\end{equation}
so to avoid a $\delta$ function in \eqref{eq:greenfcneq} we need
$c_{1}=0$. Thus $c_{2}=\frac{1}{4\pi}$.

Bringing everything together, we obtain the following expression for
the longitudinal component of the momentum perturbation at first order,
\begin{equation}
\pi^{L}(\mathbf{x})=2G_{N}\sum_{A}z_{A}^{2}\delta^{i'j'}p_{Ai'}\partial_{j'}u_{A}\left(1-\frac{2u_{A}^{2}+6u_{A}+3}{2\sqrt{u_{A}}(u_{A}+2)^{3/2}}\right)\,.\label{eq:longmom}
\end{equation}

\subsubsection{Transverse component}

Next we want to solve equation \eqref{eq:transgreen} for the transverse
Green function. Inserting the expression \eqref{eq:paralleltransp}
for the parallel transporter in terms of derivatives of $u$ leads
to the differential equation
\begin{equation}
\left(\nabla^{2}-2\right)G_{ij'}^{T}(\mathbf{x},\mathbf{x'})=\left(\partial_{i}\partial_{j'}u\right)z^{\prime\,3}\delta(\mathbf{x}-\mathbf{x'})+\left(\partial_{i}u\right)\left(\partial_{j'}u\right)\Delta_{0}''(u)+\left(\partial_{i}\partial_{j'}u\right)\Delta_{0}'(u)\,,\label{eq:transgreen3}
\end{equation}
where $\Delta_{0}(u)$ is given by \eqref{eq:inverselap1}. The $\partial_{i}u\partial_{j'}u$
term in $g_{ij'}$ vanishes in the coincident limit, so does not appear
when multiplied by the delta function. Following \citep{D_Hoker_1999}
we decompose $G_{ij'}$ into two independent tensor structures,
\begin{equation}
G_{ij'}^{T}(\mathbf{x},\mathbf{x'})=\left(\nabla_{i}\nabla_{j'}u\right)\,A(u)+\left(\nabla_{i}u\nabla_{j'}u\right)\,B(u)\,,\label{eq:tstructures}
\end{equation}
and proceed to solve for the unknown functions $A(u)$ and $B(u)$.
Using the identities in Appendix \ref{sec:appA} for various derivatives
of the chordal distance $u$ we end up with a pair of coupled ordinary
differential equations,
\begin{align}
u(u+2)A''+3(u+1)A'-A+2(u+1)B & =z^{\prime\,3}\delta(\mathbf{x}-\mathbf{x'})+\Delta_{0}'(u)\,,\label{eq:tragreen1}\\
u(u+2)B''+7(u+1)B'+2A'+2B & =\Delta_{0}''(u)\,.\label{eq:tragreen}
\end{align}
The transverse gauge condition
\begin{equation}
g^{(0)ab}\nabla_{a}G_{bj'}^{T}=0\,,\label{eq:transv}
\end{equation}
implies
\begin{equation}
3A+(u+1)A'+4(u+1)B+u(u+2)B'=0\,,\label{eq:transcond}
\end{equation}
which can be re-expressed as 
\begin{equation}
u(u+2)C'(u)+2(u+1)C(u)=2A(u)\,,\label{eq:transcond2}
\end{equation}
where we have defined the auxiliary function $C(u)$ as the following
linear combination of the transverse vector propagator functions,
\begin{equation}
C(u)=(u+1)A(u)+u(u+2)B(u)\,.\label{eq:cfunction}
\end{equation}
A suitably chosen linear combination of \eqref{eq:tragreen1} and
\eqref{eq:tragreen} yields the following ordinary differential equation
for $C(u)$ alone,
\begin{equation}
u\left(u+2\right)C''(u)+5\left(u+1\right)C'(u)=z^{\prime\,3}\delta(\mathbf{x}-\mathbf{x'})+\left(u+1\right)\Delta_{0}'(u)+u\left(u+2\right)\Delta_{0}''(u)\,.\label{eq:calone}
\end{equation}
The right hand side may be further simplified using \eqref{eq:inhomeq2}
and \eqref{eq:inverselap1} to give
\begin{equation}
u\left(u+2\right)C''(u)+5\left(u+1\right)C'(u)=\frac{u+1}{2\pi u^{3/2}(u+2)^{3/2}}\,.\label{eq:Cequation}
\end{equation}
We can solve for $C(u)$ using the same strategy as before. The homogeneous
problem is in fact identical to the one we encountered for the function
$a(u)$ when solving for the longitudinal Green function but the particular
solution to the inhomogeneous problem is different,
\begin{equation}
C(u)=c_{1}\frac{(u+1)(2u^{2}+4u-1)}{3u^{3/2}(u+2)^{3/2}}+c_{2}-\frac{u+1}{4\pi\sqrt{u(u+2)}}\,.
\end{equation}
Fall-off as $u\to\infty$ requires $\frac{2}{3}c_{1}+c_{2}=\frac{1}{4\pi}$
while at small $u$ we have 
\begin{equation}
C(u)=-\frac{1}{3(2u)^{3/2}}c_{1}+O\left(\frac{1}{\sqrt{u}}\right)\,,\label{eq:Csmallu}
\end{equation}
 which fixes $c_{1}=0$, hence $c_{2}=\frac{1}{4\pi}$. 

The next step is to solve for the transverse vector propagator functions
$A(u)$ and $B(u)$ by inserting the solution for $C(u)$ into \eqref{eq:transcond2}
and \eqref{eq:cfunction}, 
\begin{align}
A(u) & =-\frac{1}{8\pi\sqrt{u(u+2)}}\left(u+1-\sqrt{u(u+2)}\right)^{2}\,,\label{eq:Afinal}\\
B(u) & =-\frac{1}{4\pi}\left(1-\frac{(1+u)\left(u(u+2)-\frac{1}{2}\right)}{\left(u(u+2)\right)^{3/2}}\right)\,.\label{eq:Bfinal}
\end{align}
Finally, the full expression for $\pi_{i}^{T}$ is obtained by summing
over contributions from the different point particle sources,
\begin{equation}
\pi_{i}^{T}(\mathbf{x})=8\pi G_{N}\sum_{A}z_{A}^{2}\delta^{k'j'}p_{Ak'}\left(\partial_{i}\partial_{j'}u_{A}A(u_{A})+\partial_{i}u_{A}\partial_{j'}u_{A}B(u_{A})\right)\,.\label{eq:transmomentum}
\end{equation}

\section{Hamiltonian\label{sec:Hamiltonian}}

As is well-known, the Hamiltonian and momentum constraints reduce
the bulk contribution over a spacelike hypersurface to a total derivative
term, that may be expressed as a boundary contribution. Brown and
York \citep{Brown:1992br} showed that if one wishes to impose Dirichlet
boundary conditions at infinity, the term takes the form of the trace
of the extrinsic curvature of the boundary embedded in the spacelike
hypersurface. In asymptotically anti-de Sitter spacetimes, this term
can lead to extra divergent terms which may be cancelled by the addition
of boundary counterterms \citep{Henningson:1998gx,Balasubramanian:1999re}.

We begin by computing the trace of the extrinsic curvature of the
2-boundary $B$ embedded in the constant time hypersurface $\Sigma$,
using the perturbed metric \eqref{eq:metricpert} and assembling that
into the boundary Hamiltonian
\begin{equation}
H=\frac{1}{8\pi G_{N}}\int_{B}\sqrt{\det\sigma}N(k-k_{0})=\frac{1}{2\pi G_{N}}\int d^{2}x\frac{1}{z^{3}}\left(\phi+z\partial_{z}\phi+\cdots\right)\,,\label{eq:brownyork}
\end{equation}
where the subtraction term with $k_{0}=-2$ ensures that the Hamiltonian
vanishes for an unperturbed AdS background and the ... on the right
hand side denotes higher order terms in the perturbation expansion.
Noting the rapid fall-off of the trace perturbation of the metric
\eqref{eq:falloff}, we may discard the higher order contributions
$\mathcal{O}(\phi^{2})$ near the boundary, and need only retain the
$\phi+z\partial_{z}\phi$ contribution. This may be converted into
a bulk integral
\begin{equation}
\begin{aligned}H & =\frac{1}{2\pi G_{N}}\int d^{3}x\partial_{z}\left(\frac{1}{z^{3}}\left(\phi+z\partial_{z}\phi\right)\right)\\
 & =-\frac{1}{2\pi G_{N}}\int d^{3}x\sqrt{\det g^{(0)}}N^{(0)}\left(-3\phi+\nabla^{2}\phi\right)\,,
\end{aligned}
\label{eq:bulkham}
\end{equation}
with $g_{ij}^{(0)}=\delta_{ij}/z^{2}$ and $N^{(0)}=1/z$. We may
then compute the right hand side using the Hamiltonian constraint
\eqref{eq:hamconstr}, evaluated at second order. We note \eqref{eq:brownyork}
expresses the Hamiltonian as a boundary term, which can be matched
directly with the conformal field theory time evolution operator acting
on a set of boundary operator insertions. On the other hand, one also
has a bulk interpretation of the same Hamiltonian in \eqref{eq:bulkham}.
In each case, the canonical variables will be a set of particle positions
and momenta, along with the radiation degrees of freedom of the gravitational
field $\pi_{TT}^{ij}$ and $h_{ij}^{TT}$.

The resulting Hamiltonian can be expressed as a sum of three terms,
\begin{equation}
H=H_{rad}+H_{A}+H_{AB}\,,
\end{equation}
where $H_{rad}$ is the Hamiltonian of the radiation degrees of freedom,
$H_{A}$ is a set of terms dependent on the positions and momenta
of single particles (and their couplings to the radiation terms) while
$H_{AB}$ denotes terms involving pairwise couplings to other fields.
At higher order in $G_{N}$ many-body interactions will appear.

The term quadratic in the radiation fields is
\begin{equation}
\begin{aligned}H_{rad}= & -\frac{1}{16\pi G_{N}}\int d^{3}x\sqrt{g^{(0)}}N^{(0)}\bigg(g^{(0)ij}g^{(0)kl}h_{ik}^{TT}h_{jl}^{TT}-g_{ij}^{(0)}g_{kl}^{(0)}\pi_{TT}^{ik}\pi_{TT}^{jl}\\
 & \quad+g^{(0)ij}g^{(0)kl}g^{(0)mn}\left(\frac{3}{4}\nabla_{i}h_{km}^{TT}\nabla_{j}h_{ln}^{TT}-\frac{1}{2}\nabla_{i}h_{km}^{TT}\nabla_{n}h_{jl}^{TT}+h_{km}^{TT}\nabla_{i}\nabla_{j}h_{ln}^{TT}\right)\bigg)\,,
\end{aligned}
\label{eq:Hrad}
\end{equation}
while the term dependent on a single particle position/momentum takes
the form
\begin{equation}
\begin{aligned}H_{A}= & \sum_{A}\left(\frac{\bar{m}_{A}}{z_{A}}-\frac{1}{2z_{A}\bar{m}_{A}}p_{A}^{i}p_{A}^{j}h_{ij}^{TT}(\mathbf{x}_{A})\right)\\
 & \qquad-\frac{1}{8\pi G_{N}}\int d^{3}x\sqrt{g^{(0)}}N^{(0)}\left(2\nabla^{i}\nabla^{j}\phi_{1}h_{ij}^{TT}-\tilde{\pi}_{ij}\pi_{TT}^{ij}\right)\,,
\end{aligned}
\label{eq:HA}
\end{equation}
where
\begin{equation}
\bar{m}_{A}=\sqrt{g_{(0)}^{ij}p_{Ai}p_{Aj}+m_{A}^{2}}\,,\label{eq:mdefinition}
\end{equation}
and we are using the shorthand notation from \eqref{eq:shorthand}. 

The expression for $\phi_{1}$ has been given in the previous section
\eqref{eq:phi1final}, expressed as a sum over $A$, and $\nabla_{b}\nabla_{d}\phi$
may be straightforwardly computed using that formula. Likewise the
first order evaluation of $\pi_{a}^{T}$ and $\pi^{L}$ of the previous
section determines the integrand for the last term. The terms inside
the integral in $H_{A}$ reduce to total derivatives in flat spacetime,
but survive in anti-de Sitter due to the nontrivial lapse function.

The final term to be assembled is $H_{AB}$. Here we run into the
issue that terms arising from sources where $B=A$ give rise to divergent
terms. In the flat spacetime case \citep{Schafer:1986aa} this type
of term is dealt with via Hadamard's method of partie finie \citep{hadamard}.
Instead we note these terms are divergent and independent of the particle
position. Therefore they may be removed by position independent counterterms
in the Hamiltonian via renormalization. We proceed with the understanding
that only the finite $B\neq A$ terms are to be considered. With that
in mind,
\begin{equation}
\begin{aligned}H_{AB}= & \sum_{A,B\neq A}\frac{2}{z_{A}\bar{m}_{A}}\phi_{1B}(\mathbf{x}_{A})g_{(0)}^{(0)ij}(x_{A})p_{Ai}p_{Aj}\\
 & \quad-\frac{1}{16\pi G_{N}}\int d^{3}x\sqrt{g^{(0)}}N^{(0)}\left(96\phi_{1}^{2}-16\phi_{1}\nabla^{2}\phi_{1}-8(\nabla\phi_{1})^{2}-\tilde{\pi}^{ij}\tilde{\pi}_{ij}\right)\,.
\end{aligned}
\label{eq:HAB}
\end{equation}
 As before, the integrand in this expression is given explicitly in
terms of the expression \eqref{eq:phi1final} for $\phi_{1}$ and
the first order evaluation of of $\pi_{a}^{T}$ and $\pi^{L}$ of
the previous section. 

This concludes the derivation of the Hamiltonian at linear order in
$G_{N}$ in the post-anti-de Sitter approximation, where the dynamical
variables are the set of particle momenta and positions $(x_{A},p_{A})$
and the transverse-traceless radiation variables $(h_{ij},\pi_{TT}^{ij})$.
The answer contains integrals that resemble one-loop integrals in
quantum field theory in anti-de Sitter spacetime. It seems likely
some or all of these integrals can be obtained in closed form, however
we leave that issue for future work. We have checked that the integrals
fall off as the separation of the particles increases, and vanish
in the limit that a particle approaches the boundary. In the flat
spacetime case, the corresponding integrals can be explicitly performed
and are tabulated in appendix 3 of \citep{Ohta:1974pq}. 

It should be noted the conformally flat metric we have chosen does
not provide a set of global coordinates on AdS. Instead one must glue
together a sequence of such coordinates patches to cover the (universal
cover of) AdS. To fully explore this, one must then provide matching
conditions for the particle momenta and positions as they transition
from one coordinate patch to another. We will not attempt a detailed
construction here.

\section{Conclusions}

In this work we have obtained the Hamiltonian for $N$-particles of
arbitrary masses undergoing mutual gravitational interactions, at
leading order in $G_{N}$ and to all orders in momenta, akin to the
so-called post-Minkowskian approximation to general relativity in
asymptotically flat spacetime \citep{Schafer:1986aa}. At this order,
a pairwise interaction is present, in addition to couplings to gravitational
waves. In AdS spacetime, generic particles remain at finite separation
on average, so interactions will make finite contributions to time-averaged
observables. This is rather different from a generic scattering process
in asymptotically flat spacetime where the particles scatter off each
other and move off to infinity. However, for special initial conditions
in asymptotically flat spacetime that correspond to mutually bound
orbits, the persistent pairwise interactions lead to chaotic behavior
\citep{Portegies_Zwart_2022}. 

This provides strong evidence that there is a hard upper bound on
the Lyapunov time governing typical scattering states in a conformal
field theory dual to gravity in an asymptotically anti-de Sitter background.
 We expect this upper bound to change qualitatively once one goes
to higher orders, due to the 3-body and beyond interactions that then
begin to appear. However this will only serve to lower the bound.
As argued in the introduction, this in turn provides strong evidence
that the spectrum of primary operators in the CFT will exhibit level
repulsion, in line with the Wigner surmise.
\begin{acknowledgments}
We thank G. Schafer for helpful comments. The research of D.L. is
supported in part by DOE grant de-sc0010010 and that of L.T. in part
by the Icelandic Research Fund Grant 228952-053 and by the University
of Iceland Research Fund. We would like to thank the Isaac Newton
Institute for Mathematical Sciences, Cambridge, for support and hospitality
during the program \textquotedblleft Black Holes: Bridges Between
Number Theory and Holographic Quantum Information\textquotedblright{}
where work on this paper was undertaken. This work was supported by
EPSRC grant no EP/K032208/1.
\end{acknowledgments}

\appendix

\section{Useful identities\label{sec:appA}}

In this appendix we have collected together some identities satisfied
by the chordal distance variable \eqref{eq:chordaldistance} and its
derivatives that are referred to in the main text. 

\begin{align}
g^{(0)ij}\left(\nabla_{i}u\right)\left(\nabla_{j}u\right) & =u(u+2)\,,\label{eq:id1}\\
\nabla_{i}\nabla_{j}u & =g_{ij}^{(0)}(u+1)\,,\label{eq:id2}\\
g^{(0)ij}\left(\nabla_{i}u\right)\left(\nabla_{j}\nabla_{j'}u\right) & =(u+1)\nabla_{j'}u\,,\label{eq:id3}\\
\nabla_{i}\nabla_{j}\nabla_{j'}u & =g_{ij}^{(0)}\nabla_{j'}u\,,\label{eq:id4}\\
\nabla^{2}\left(\nabla_{i}\nabla_{j'}u\right) & =\nabla_{i}\nabla_{j'}u\,,\label{eq:id5}\\
\nabla^{2}\left(\nabla_{i}u\nabla_{j'}u\right) & =4\nabla_{i}u\nabla_{j'}u+2(u+1)\nabla_{i}\nabla_{j'}u\,,\label{eq:id6}\\
g^{(0)ab}\left(\nabla_{b}u\right)\left(\nabla_{a}\nabla_{i}\nabla_{j'}u\right) & =\nabla_{i}u\nabla_{j'}u\,,\label{eq:id7}\\
g^{(0)ab}\nabla_{a}\left(\nabla_{i}u\nabla_{j'}u\right)\nabla_{b}u & =2(1+u)\nabla_{i}u\nabla_{j'}u\,.\label{eq:id8}
\end{align}

\bibliographystyle{utcaps}
\bibliography{cft_chaos}

\end{document}